\documentclass[%
 reprint,
 amsmath,amssymb,
 aps,
pra,
]{revtex4-2}

\usepackage{graphicx}
\usepackage{dcolumn}
\usepackage{bm}
\usepackage{tikz}
\usepackage{tikz-feynman}
\usepackage{hyperref}
\usepackage{subfigure}

\begin{document}

\preprint{APS/123-QED}

\title{Fractional quantum Hall states by Feynman's diagrammatic expansion} 

\author{Ben Currie}
\author{Evgeny Kozik}%
\affiliation{%
 Department of Physics, King’s College London, Strand, London WC2R 2LS, United Kingdom
}%

\date{\today}

\begin{abstract}

The fractional quantum Hall (FQH) effect arises from strong electron correlations in a quantising magnetic field, and features exotic emergent phenomena such as electron fractionalisation. Using the diagrammatic Monte Carlo approach with the combinatorial summation (CoS) algorithm, we obtain results with controlled accuracy for the microscopic model of interacting electrons in the lowest Landau level (LLL) in the thermodynamic limit. \textcolor{black}{Starting from the macroscopically degenerate LLL at finite temperature, including interactions order by order, and applying a controlled resummation to the resulting series, we observe the emergence of the incompressible 1/3-filled state as the temperature is lowered. }
By analysing the long-time decay of the Green's function, we find spectral properties consistent with an energy gap at $1/3$-filling, whereas at $1/2$-filling our results are consistent with the pseudogapped behaviour previously observed experimentally and suggested theoretically. 
Our work provides the first demonstration that fractionalised phases of matter can be reliably described with Feynman's diagrammatic technique in terms of the fundamental electronic degrees of freedom, while also showing applicability of expansions in the bare Coulomb potential for precision calculations.

\end{abstract}

\maketitle

The application of a strong magnetic field to a two-dimensional electron gas leads to a rich variety of quantum phases. The resulting formation of perfectly flat bands -- the well-known Landau levels-- provides an ideal environment for electronic correlations at partial band filling, where Coulomb interactions dominate. This is perhaps evidenced best by the emergence of the fractional quantum Hall (FQH) states \cite{Tsui1982,Stormer1983}, which are prototypical examples of topologically ordered phases of matter and which play host to low-energy excitations with fractional charge and statistics \cite{Arovas1984,Halperin1984}. While a remarkable amount of the Landau-level physics is described extremely well in terms of composite Fermions \cite{Jain1989,Jain1990,Lopez1991,Zhang1989,HLR1993,Shankar1997}, including not only the FQH states but also the composite Fermi liquids at even denominator fillings, an understanding of the effect based solely on the electronic degrees of freedom has thus far not been achieved. Moreover, very little is known about the possible finite-temperature phases.

The lack of kinetic energy implies that the physics of partially filled Landau levels is inherently driven by strong electron correlations, which is challenging to study due to the absence of a small parameter. This has thus far prevented the development of controlled calculational schemes to study the microscopic electronic theory in the thermodynamic limit. Instead, controlled numerical solutions have come only from intrinsically finite-size methods such as exact diagonalisation \cite{YHL1983,Laughlin1983b,Haldane1983,Haldane1985,Fano1986,d'Ambrunmenil1989,Morf2002} and DMRG \cite{Naokazu2001,Feiguin2008,Kovrizhin2010,Zaletel2012,Hu2012,Estienne2013,Zaletel2013}. Moreover, effective field theories based on composite Fermions provide qualitatively correct physics at mean-field level, but the systematic improvement by incorporating higher-order diagrams is hindered by the complexity of the perturbative expansion, which involves expanding in an emergent gauge field.

In this paper, we study the microscopic theory of electrons in the partially filled lowest Landau level (LLL) in terms of the bare electrons and include the effect of interactions order-by-order by an exact diagrammatic expansion around the non-interacting limit. Owing to the macroscopic degeneracy of the partially filled LLL, one might object that such a perturbation theory is not well-defined, due to the complete restructuring of the state when arbitrarily weak interactions are turned on: a consequence of the fact that the interaction provides the only energy scale $V$. However, the introduction of a finite temperature $T$ supplies an additional energy scale, and thereby also provides a perturbative regime in which $V/T \ll 1$. A qualitative change will occur as we pass from this regime to the strong coupling limit $V/T\gg 1$, in which the FQH states are expected to emerge via a smooth crossover.
This is similar to the viewpoint exploited in controlled diagrammatic calculations for a state qualitatively distinct from the non-interacting limit, for example, in the metal-to-insulator crossover in the $2d$ Hubbard model~\cite{Simkovic2020crossover, Kim2020spin_charge}.
However,
previous approximate diagrammatic approaches to the LLL in terms of the fundamental electronic degrees of freedom have not found a FQH state \cite{TaoThouless1983,Thouless1985,Haussmann1996, Currie2024}. Other results in the thermodynamic limit based on a high-temperature expansion were restricted to temperatures much larger than the gap \cite{Zheng1994,Sawatdiaree2000}. A previous study by the present authors~\cite{Currie2024} based on an exact resummation of a specific subset of diagrams, corresponding to the self-consistent $GW$ approximation, gave a non-Fermi liquid state in a broad range of filling fractions, but no evidence for FQH states was seen.

Here we apply a formally exact Feynman-diagrammatic expansion in the bare Coulomb potential with variable screening and evaluate the series numerically exactly using the diagrammatic Monte Carlo (DiagMC) technique~\cite{Prokofev1998, Prokofev2007, VanHoucke2010, Kozik2010} with the deterministic combinatorial summation (CoS) of all diagram integrands~\cite{Kozik2024}. \textcolor{black}{ We address both the short- and long-range screening regimes (relative to $\ell_B$) and show that this enables an extrapolation to the pure Coulomb limit.} The flexibility of the CoS framework allows us to significantly improve the convergence properties of the series by compensating all diagrams that contain insertions contributing to the exact density with an appropriate shift of the chemical potential~\cite{VanHoucke2010, Wu2017}. Through calculating the expansion coefficients with high accuracy up to a sufficiently high diagram order ($n=8$ in practice), we can control the precision of reconstructing the results from the series in regimes where it proves divergent \textcolor{black}{and FQH physics is expected to appear}. 

We demonstrate the emergence of the FQH state at $1/3$ filling of the LLL and probe its development as the temperature is lowered by studying the density against chemical potential (the equation of state): We observe a prominent plateau at $1/3$ developing at low temperature, together with a compressibility saturating to a non-zero value at half-filling. Correspondingly, a gap opens in the single-electron spectral function $\rho(\omega)$ at $1/3$-filling, while the spectral function of the $1/2$ state \textcolor{black}{exhibits a pseudo-gap at a much lower temperature scale, consistent with experiments and numerical calculations}. \textcolor{black}{ These features are qualitatively the same for all studied interaction ranges including for the pure Coulomb potential.} Our results demonstrate that the emergence of an FQH state can be probed by a Feynman diagrammatic expansion method relying on an extrapolation of a large but ultimately finite number of diagram orders. They also show that, despite their intrinsic divergence, bare expansions in the Coulomb potential are a viable approach to many-electron systems, which opens the door to more tractable simulations of real materials.

We study the continuum model of electrons in two dimensions subject to a uniform and perpendicular magnetic field with strength $B$ and with density-density interactions.
The corresponding Hamiltonian is 
\begin{equation} \label{LLL Hamiltonian}
    H = \sum_{i}\frac{1}{2m}(-i\partial_{\boldsymbol{r}_i} +e\boldsymbol{A}(\boldsymbol{r}_i))^2 + \frac{\xi}{2}\sum_{i,j} V(\boldsymbol{r}_i-\boldsymbol{r}_j) ,
\end{equation}
where $\boldsymbol{A}$ is the vector potential corresponding to the external magnetic field, $V(\boldsymbol{r}_i-\boldsymbol{r}_j)$ is the Coulomb potential between the electrons at the $2d$-vector coordinates $\boldsymbol{r}_{i,j}$ and $\xi$ is an auxiliary expansion parameter, which can be viewed as a (dimensionless) coupling constant set to unity at the end of the calculation.
In the non-interacting limit ($\xi = 0$), the single-particle spectrum reduces to the familiar Landau levels, which have a degeneracy per unit area $\mathcal{D} = 1/2\pi \ell_B^2$ and are spaced in energy by $\hbar\omega_B$, where $\omega_B = eB/m$ is the cyclotron frequency and $\ell_B = \sqrt{\hbar/eB}$ is the magnetic length.
For simplicity, we consider the large-$B$ limit with $\omega_B \gg e^2/\ell_B$, such that all electrons occupy the lowest Landau level, and all higher levels are projected out; generalisations are straightforward. 
 
Our observable of interest is the fully interacting finite-temperature Matsubara Green's function projected onto the LLL, defined as \cite{AGD}
\begin{equation}
    G(\boldsymbol{r}_1 \boldsymbol{r}_2,\tau) = -\langle \mathcal{T}\tilde\phi(\boldsymbol{r}_1,\tau) \tilde\phi^\dagger(\boldsymbol{r}_2,0)\rangle,
\end{equation}
where $\tilde\phi$ is the LLL-projected electron annihilation operator and $\mathcal{T}$ is the time-ordering operator for the imaginary time $\tau$. Remarkably, provided the translational symmetry is unbroken, the spatial dependence of $G$ is independent of the interaction strength, and is given by
\begin{equation} \label{GF}
    G( \boldsymbol{r}_1 \boldsymbol{r}_2, \tau) = \frac{1}{2\pi\ell_B^2}G(\tau)e^{-\frac{1}{4 \ell_B^2}| \boldsymbol{r}_1-\boldsymbol{r}_2|^2}e^{i\varphi_{\boldsymbol{r}_1, \boldsymbol{r}_2}},
\end{equation}
where 
$    \varphi_{ \boldsymbol{r}_1, \boldsymbol{r}_2} = \frac{2\pi}{\phi_0}\int_{ \boldsymbol{r}_1}^{ \boldsymbol{r}_2} \boldsymbol{A}(\boldsymbol{x}) \cdot d \boldsymbol{x}$
is the Aharonov-Bohm phase picked up by an electron traversing the straight-line path from $\boldsymbol{r}_1$ to $\boldsymbol{r}_2$, 
and where
$\phi_0 = h/e$ is the flux quantum. Here,  $G(\tau)$ is the (gauge-invariant) Green's function in the basis of LLL states, whose precise form (at a given filling and temperature) depends on the interaction strength and encodes the single-electron excitation properties of the system. 

Diagrammatic expansions in the bare Coulomb potential are generally avoided due to two notorious issues: (i) the Dyson collapse argument~\cite{Dyson1952} implies that such an expansion in continuous space has zero convergence radius and (ii) the long-range nature of the potential makes certain individual diagrams diverge upon integration over spatial coordinates. The zero convergence radius is a result of the discontinuous change of the state of the system, a collapse to infinite density, when the sign of the coupling constant $\xi$ is flipped. Just as in lattice problems, it does not apply here because the density is bounded by the capacity of the (flat) LLL. The second problem is typically dealt with by screening the Coulomb interaction with dynamical polarisation and expanding in the screened coupling~\cite{Hedin1965}, which eliminates the divergent diagrams to avoid double counting. In another approach~\cite{Chen2019}, one introduces an auxiliary Yukawa screening, which renders the integrals convergent and can be tuned to accelerate the series convergence, and compensates the artificial screening by counter terms. 

\begin{figure}
\centering
\includegraphics[trim = {9 0 -9 0},clip,width=0.5\textwidth]{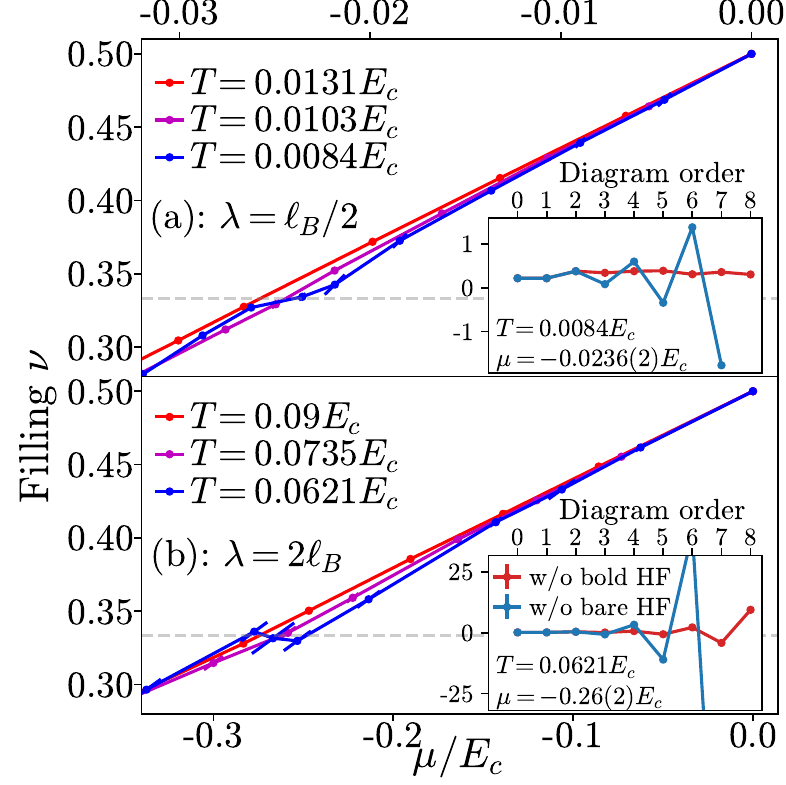}
\caption{The filling against chemical potential (equation of state) for the lowest Landau level \eqref{LLL Hamiltonian} with Yukawa potential (defined in the main text) and screening parameter $\lambda = \ell_B/2$ (a) and $\lambda = 2 \ell_B$ (b) at different temperatures. Chemical potential $\mu$ is defined relative to the half filling value, and $E_c = e^2/\ell_B$. The insets show the partial sums of the series at low temperature and at chemical potential on the plateau. The series without bold Hartree and Fock (w/o bold HF) diagrams shows substantially better convergence than the series in which only the \textit{bare} Hartree and Fock diagrams are excluded (w/o bare HF). 
}
\label{fig:EOS}
\end{figure}

In contrast, we employ the Yukawa potential $V(r) =  e^{-r/\lambda}/r$ as the physical interaction and focus on the specific values $\lambda =2 \ell_B$ and $\lambda =\ell_B/2$ such that both the long-range Coulomb ($\lambda>\ell_B$) and short-range  ($\lambda < \ell_B$) regimes are captured. The short-range regime is \emph{a priori} a simpler case for observing FQH physics, is relevant in the context of model studies \cite{Haldane1983,Haldane1985,Trugman1985,Pokrovsky1985} \textcolor{black}{as well as for real systems with gate-screening \cite{Aleiner1994}}, and naturally results in a better-behaved diagrammatic series. 
The regime of long-range interaction is of fundamental importance, both for the question of its effect on the possible FQH state and for studying the applicability of expansions in the Coulomb potential.  In the pure Coulomb limit $\lambda \to \infty$, all non-skeleton in the coupling channel diagrams---those that can be made disconnected by cutting any two interaction lines---diverge upon integration over internal variables. Thus, at large $\lambda$ the series will be dominated by these terms, the role of which is to dress all interaction lines in all other diagrams by electron polarisation screening. A simple analysis (see Appendix~\ref{Yukawa appendix}) shows that the divergence is due to a singularity at location $\xi_s \sim -\lambda^{-1}$. Analytic continuation beyond the radius of convergence $|\xi_s|$ thus gives access to the Coulomb regime, in which the electron gas self-screens the long-range tail of the potential and thus all physical quantities are subject only to weak (\textcolor{black}{linear}) corrections in $\lambda^{-1}$, which can be extrapolated away. \textcolor{black}{We show such an extrapolation in the supplementary material.}

We compute $G(\tau)$ directly in the thermodynamic limit and at finite temperature $T$ in the DiagMC framework~\cite{Prokofev1998, Prokofev2007, VanHoucke2010, Kozik2010}. In this approach, an arbitrary physical observable $A$ is obtained---without approximations---from its Taylor series $\sum_{n=0}^N a_n \xi^n$ in the parameter $\xi$ up to some maximal diagram order $N$, whereby $a_n$ is represented as a sum over all diagrams with $n$ interaction lines. Within each order $n$, all diagram integrands are summed deterministically by the CoS algorithm~\cite{Kozik2024} and integrated over the positions of the interaction vertices in space-imaginary-time by stochastic sampling, which gives the exact value of $a_n$ with a known statistical error. We use the version of the algorithm with symmetrisation over all vertex permutations, which takes $\mathcal{O}(n^3 3^n)$ floating point operations to sum all connected diagram topologies of order $n$ for spinless fermions~\cite{Kozik2024}. The extra computational cost compared to $\mathcal{O}(n^2 2^n)$ for the series with ordered vertices is compensated by the much smaller Monte Carlo variance for the symmetrised integrand.

To improve the convergence properties of the series~\cite{VanHoucke2010, Wu2017}, we formulate the expansion in terms of the propagator partially dressed by the bold Hartree-Fock self-energy, as in the original DiagMC scheme~\cite{VanHoucke2010}: 
\begin{equation} \label{HF GF}
    \tilde{G}_0(i\omega) = (i\omega+\mu-\Sigma^{\mathrm{bold}}_{\mathrm{HF}})^{-1},
\end{equation}
where $\mu$ is the chemical potential, $i\omega$ is a fermionic Matsubara frequency, $\Sigma^\mathrm{bold}_{\mathrm{HF}} = V_{\mathrm{HF}} \nu$ with $\nu$ the exact filling fraction and $V_\mathrm{HF} = (1/2\pi\ell_B^2) \times\int d^2\boldsymbol{r}~ V(\boldsymbol{r})(1 - e^{-\boldsymbol{r}^2/2\ell_B^2})$ is a constant equal to the uniform component of the Hartree-Fock effective potential. 
To avoid double-counting, all diagrams containing an insertion that is part of $\Sigma^{\mathrm{bold}}_{\mathrm{HF}}$ must be excluded in the determination of $a_n$. This is straightforward to implement within the CoS approach by forbidding such diagrams during the construction of the computational graph~\cite{Kozik2024}; the resulting increase in computational cost is insignificant for the expansion orders considered.
\textcolor{black}{In the insets of Fig. \ref{fig:EOS}, we compare this series to one where only bare Hartree-Fock diagrams are excluded; the former displays a substantially enlarged convergence radius.}

We extrapolate the series to infinite diagram order using  Pad\'e and Dlog-Pad\'e resummation \cite{Baker1961}, whereby a sequence of approximants are constructed based on the Taylor series coefficients alone, and are used to reconstruct the function at the physical coupling $\xi = 1$. The discrepancy between different approximants with arbitrary free parameters yields the total systematic error of the extrapolation, as explained in Ref.~\cite{Simkovic2019} and detailed in appendix \ref{Pade appendix}.  \textcolor{black}{As the temperature is lowered the series becomes increasingly divergent, which leads to an increased but estimable systematic error of the extrapolation. Therefore, the lowest accessible temperature is limited by the largest diagram order we can compute with reasonable statistical error.}

The dependence of the filling fraction $\nu$, obtained from the Green's function via $\nu = -G(\beta^{-})$, on the chemical potential is shown in Fig.\ref{fig:EOS}a for $\lambda =\ell_B/2$ and for $\lambda = 2\ell_B$ in Fig.\ref{fig:EOS}b.
Due to the symmetry $\nu \to 1-\nu$ for $\mu \to -\mu$, only negative chemical potentials are shown.  At high temperatures, where the series converges quickly, the equation of state is featureless, and the system is compressible for all fillings shown. However, as the temperature is lowered, a clear plateau is formed around the filling  $\nu =1/3$, indicative of the opening of a gap and the formation of the $1/3$ FQH state. The size of the plateau provides a rough estimate for the gap size, which after accounting approximately for thermal broadening effects is consistent with the values $\Delta(\lambda =\ell_B/2) \sim 0.01 e^2/\ell_B$ and $\Delta(\lambda=2\ell_B) \sim 0.07e^2/\ell_B$. The reduced gap in the smaller screening $\lambda$ case is due to the weakening of the interaction energy scale by the reduced potential range and is consistent with the similar rescaling of the corresponding $T$ and $\mu$ . In the $\lambda \to 0$ limit, this reduction is by the factor $\lambda^3$. Hence, for small $\lambda < \ell_B$ we can account approximately for the reduction by dividing all energy scales by $\lambda^3$, which, for both $\lambda$ values, gives a gap size in qualitative agreement with estimates based on the composite-fermion theory~\cite{Jain1997}. 
On the other hand, at half-filling, the system remains compressible down to the lowest temperatures, with an apparent saturation of the compressibility to the zero temperature regime.  \textcolor{black}{For reference, the Coulomb energy scale corresponds to $e^2/\ell_B \approx 180 ~\mathrm{K}$ for experiments in GaAs at a typical magnetic field $B = 13 ~\mathrm{T}$.}

Mathematically, the emergence of the plateau at $\nu=1/3$ appears to be due to a particular structure of singularities in the Taylor series for density or Green's function. As $\tilde{\mu}$ is increased towards $\nu=1/3$, the levelling of the curve coincides with a rapid movement of a complex-conjugate pair of apparent singularities with a large positive real part into the circle $|\xi|=1$. A further increase of $\tilde{\mu}$ towards half-filling ($\tilde{\mu}=0$) does not shift them significantly. However, a sequence of plateaus at odd denominator fractions is expected to emerge with further cooling (beyond that in Fig.~\ref{fig:EOS}), with the next one at $\nu = 2/5$, since its gap is estimated at a half the $\nu = 1/3$ gap \cite{Jain1997}. We thus conjecture that the emergence of the $\nu = 2/5$ state is caused by an additional pair of singularities approaching $|\xi|=1$ from the positive side of the real axis on top of the structure defined by the $1/3$ state in the range $1/3 < \nu \leq 1/2$, and to capture it reliably evaluation of the series to a higher order with similar error bars ($\lesssim 1\%$) would be required. 

\textcolor{black}{The electron spectral function $\rho(\omega)$ has been probed experimentally by measurements of the tunnelling current between two parallel $2d$ electron gas systems in a large magnetic field \cite{Ashoori1990,Ashoori1993,Eisenstein1992,Brown1994,Eisenstein1995,Eisenstein2016}. At low temperatures, they observe a strong suppression in the current, and thus also in $\rho$, in a wide range of the LLL fillings including $\nu=1/3$ and $\nu=1/2$; i.e. regardless of whether the state is gapped or gapless. At $\nu=1/3$, such a suppression results naturally from the gap, but it's origin is less clear for $\nu=1/2$. Subsequently, a consistent theoretical picture was provided using the composite fermion approach \cite{HePlatzmanHalperin1993,Kim1994,Levitov1997,Yue2024}, leading to an exponential low-frequency dependence of the form $\rho(\omega\to 0) =  e^{-\omega_0/\omega}$, where $\omega_0$ is a roughly filling-independent energy scale on the order of $ 0.5 e^2/\ell_B$. 
In our approach, a key diagnostic into the spectral properties is offered by the decay of the Green's function at long imaginary times: 
It follows from the spectral representation that $\rho(\omega)$ near $\omega=0$ can be probed by examination of the quantity $\tilde\rho = -\beta G(\tau=\beta/2)$ \cite{WenWang2020}. }

\begin{figure}
    \centering
    \includegraphics[trim={15 0 0 0},clip,width=\linewidth]{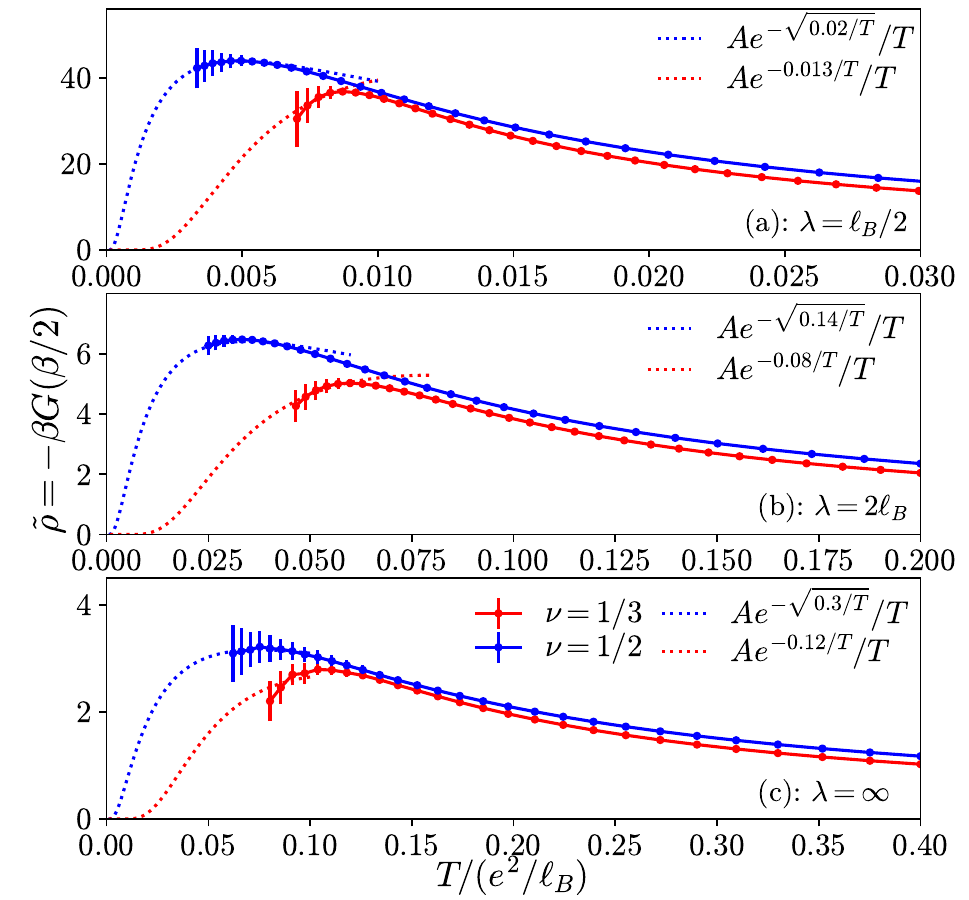}
    \caption{The Green's function at imaginary time $\tau = \beta/2$ against temperature for screening parameters (a) $\lambda = \ell_B/2$, (b) $\lambda = 2\ell_B$, and (c) $\lambda = \infty$ obtained via extrapolation. 
    Curves are at constant Hartree-Fock-shifted chemical potential, which at half-filling corresponds to constant physical chemical potential $\tilde\mu = \mu = 0$ (blue curves). The curve labelled $\nu\approx1/3$ has a filling that varies from $\nu(T\to \infty) = 0.22$ to $\nu(T\to 0) \approx 1/3$. The curves are asymptotic at high temperature to $\lim_{T \to \infty} G(\beta/2)  = -\sqrt{\nu(1-\nu)}$. For all curves, the crossover from the high-temperature region, in which the Green's function remains close to its asymptote, to the low-temperature regime occurs at the temperature scales $T \approx 0.01 e^2/\ell_B$ for $\lambda =\ell_B/2$ and $T\approx 0.06 e^2/\ell_B$ for $\lambda = 2\ell_B$, consistent with those at which the plateaus emerge in Fig\ref{fig:EOS}.}
    \label{fig:beta_over_2}
\end{figure}

\textcolor{black}{The temperature dependence of $\tilde\rho$ is shown in Fig.~\ref{fig:beta_over_2}. Here we additionally show data obtained for the Coulomb potential ($\lambda=\infty$) using a controlled extrapolation from screening ranges $\lambda >\ell_B$ (see Appendix for details). At the lowest temperatures and for all values of the Yukawa screening, our data for $\tilde\rho$ at $\nu = 1/3$ shows an exponential suppression with cooling, consistent with the opening of a gap in $\rho(\omega)$ at a temperature at which the plateau emerges in the equation of state (Fig.\ref{fig:EOS}a). It is captured by $G(\tau) \sim \exp(-\tilde{\Delta}_{1/3} \tau)$ with a decay constant $\tilde{\Delta}_{1/3}$, signalling the emergence of the gap on the order of $\tilde{\Delta}_{1/3}$, which is consistent with the sizes of the plateaus for both values of $\lambda$ shown in Fig.\ref{fig:EOS}. For the pure Coulomb interaction ($\lambda=\infty$), our estimate of $\tilde{\Delta}_{1/3}$ is consistent with recently obtained data for the local density of states within the composite-fermion framework \cite{Pu2024}.}

\textcolor{black}{The strong decay of $\tilde\rho$ at $1/3$ filling is to be contrasted with the behaviour at $\nu = 1/2$: a much weaker suppression develops gradually at a significantly smaller temperature. This suppression is present despite the fact that, in Fig.\ref{fig:EOS}, the compressibility is finite and appears temperature independent below $T \approx 0.01 e^2/\ell_B$ for $\lambda =\ell_B/2$ and $T \approx 0.08e^2/\ell_B$ for $\lambda = 2\ell_B$.}
\textcolor{black}{The low-temperature asymptote of $\tilde\rho(T)$ at $\nu=1/2$ is consistent with the pseudo-gapped spectral function $\rho(\omega\to 0) =  e^{-\omega_0/\omega}$, which corresponds to an imaginary-time dependence $G \sim \exp({-2 \sqrt{\omega_0\tau}})$ (fits of the resulting $\tilde\rho(T)$ are shown in Fig.\ref{fig:beta_over_2}). Due to appreciable systematic uncertainties in this regime, only a slightly worse fit can be achieved assuming a gapped state with a much smaller energy scale. 
Experimentally, the highest temperature at which the pseudo-gap behaviour is visible at half-filling is $T=9.5~ \mathrm{K}$ \cite{Eisenstein1992}. 
At magnetic field $B = 13 \mathrm{T}$, after taking into account the $\lambda$-dependent rescaling of the Coulomb energy, this temperature corresponds to  $0.006 e^2\ell_B$ for $\lambda = \ell_B/2$  and $0.05 e^2/\ell_B$ for $\lambda = 2\ell_B$, which both are approximately where we observe the crossover to the pseudogap in Fig.\ref{fig:beta_over_2}.}

In summary, we have studied the minimal microscopic model of FQH physics and demonstrated that the development of the $1/3$-filled Laughlin state can be obtained with controlled accuracy via the standard diagrammatic expansion around the non-interacting limit at finite temperatures. A low-temperature plateau in the equation of state emerges at $1/3$-filling due to a particular analytic structure of the resulting Taylor series and despite the absence of any prior information about it in the microscopic model or the construction of the diagrammatic expansion. This is fundamentally different to the DiagMC solution for the insulating regime of the $2d$ Hubbard model~\cite{Simkovic2020crossover, Kim2020spin_charge}, where the opening of the gap at half-filling is predetermined by the particle-hole symmetry and the ensuing nesting of the Fermi surface. In contrast, our solution gives a compressible state at half-filling that nevertheless features a pseudo-gapped spectrum at the lowest temperatures which is consistent with experiments \cite{Eisenstein1992}. This is also highly non-trivial because the low-order solution for $\nu=1/2$ is actually insulating at a relatively high temperature $T \sim 0.1 e^2/\ell_B$, and the low temperature pseudo-gapped properties are only recovered at higher orders.    

Our work provides a new window for studying the emergence of topologically non-trivial phases as a function of temperature, and is the first demonstration that a phase of matter featuring fractionalisation can be obtained in a diagrammatic expansion in terms of the prime electronic degrees of freedom. While we have focussed only on the lowest Landau level here, the approach can be readily applied to include any number of 
Landau levels and could provide a novel probe for higher-filling states, such as those thought to possess non-Abelian anyonic excitations~\cite{Willett2013Oct} and which are promising potential platforms for fault-tolerant topological quantum computation~\cite{Nayak2008}.

More generally, this shows that systems with completely flat bands can be reliably studied by finite-temperature diagrammatic expansions in the bare coupling, which \textit{a priori} may seem counter-intuitive because the absence of kinetic energy appears to make a perturbation theory impossible. In fact, since the temperature is the only energy scale for the interaction, the expansion in the powers of $\xi$ is equivalent to expanding in $V/T$ about $V/T=0$, which is a well-defined starting point. Provided the low-temperature state is not separated from this starting point by a phase transition, it is natural for a Taylor series in $\xi V/T$ to be able to capture its properties with enough expansion coefficients at hand. An advantage of the diagrammatic formulation of such a high-temperature expansion is that there are many established tools for accelerating it~\cite{homotopic_action}, as we do here by a particular shift of the chemical potential.

One of the main difficulties in simulating real materials in the diagrammatic framework~\cite{Hedin1965, Aryasetiawan_1998} is the derivation and parametrisation of the dynamically screened interaction $W$~\cite{Leon2021GW}. The Coulomb interaction $V$, in contrast, is static and straightforward to obtain in any basis. Our results suggest that diagrammatic expansions in $V$ with a regularisation of the integrals by $\lambda \to \infty$ can be used for controlled calculations when the number of computed expansion coefficients is sufficient for a reliable resummation of the series beyond the convergence radius $\sim |\lambda^{-1}|$. Since the limiting singularity is of a very simple general form (see Appendix \ref{Yukawa appendix}), the analytic continuation is robust already with a few diagram orders, which makes expansions in the bare Coulomb potential evaluated by DiagMC a promising route to controlled simulations of materials.

\begin{acknowledgments}
This work was supported by EPSRC through Grant No. EP/X01245X/1. The calculations were performed using King's Computational Research, Engineering and Technology Environment (CREATE). This work used the ARCHER2 UK National Supercomputing Service (https://www.archer2.ac.uk) \cite{ARCHER}.
\end{acknowledgments}

\bibliography{biblio}

\newpage
\onecolumngrid
\noindent
\centerline{\textbf{\large End-Matter}}
\appendix
\section{Regularisation of the bare series by Yukawa screening} \label{Yukawa appendix}
Within the bare series, all diagrams with insertions of the electron polarisation $\Pi$ in any of the interaction lines are divergent for the pure Coulomb potential ($\lambda =\infty$) due to integration of the internal vertices over space (or momentum). Let us regularise the integrals by finite $\lambda$ and ask: given the divergence of these diagrams at $\lambda =\infty$, how can we understand the limit $\lambda \to \infty$ from the bare series? Consider, for example, the $n$-th order contribution to the screened interaction $W$, which will be dominated by the diagram containing the maximal number ($n-1$) of polarisation bubbles $\Pi_0$. In the basis of LLL states, this diagram is (in units with $\ell_B=1$; see Ref.\cite{Currie2024} for details)
\begin{equation}
   W_n(i\nu) =  \int_0^\infty d\rho~ \rho ~ \xi U(\rho) (\xi\Pi_0(i\nu) U(\rho))^{n-1} ,
   \label{W_n}
\end{equation}
where $\Pi_0(\tau) = G(\tau)G(-\tau)$ is the bare electron polarisation, $G(\tau)$ is the Green's function in the LLL basis defined in Eq.~\eqref{GF}, $U(\rho) = e^{-\frac{1}{2}\rho^2}/\sqrt{\rho^2+\lambda^{-2}}$ is the LLL-projected Yukawa-screened Coulomb potential in momentum representation, and $i\nu$ is a bosonic Matsubara frequency. In the limit $\lambda \to \infty$, this diagram diverges as $\log\lambda^{-1}$ for $n=2$ and as $\lambda^{n-2}$ for $n > 2$. In the former case, performing the integration over momentum yields
\begin{equation}
\begin{split}
   W_2(i\nu) &=  \Pi_0(i\nu) \int_0^\infty d\rho ~\rho \left(\frac{\xi}{\sqrt{\rho^2+\lambda^{-2}}}e^{-\frac{1}{2}\rho^2}\right)^{2} \\&= -\xi^2\Pi_0(i\nu) \big[\gamma/2+\log(\lambda^{-1}) \big] +\mathcal{O}(\lambda^{-2}) ,
\end{split}
\end{equation}
where $\gamma$ is the Euler-Mascheroni constant.
For $n>2$, due to the domination of the small-$\rho$ region at $\lambda^{-1} \to 0$, we can ignore the Gaussian factor, which yields
\begin{equation}
\begin{split}
    W_{n>2}(i\nu) &=  \xi^n\Pi_0(i\nu)^{n-1}\int_0^\infty  d\rho~\rho \left(\frac{1}{\sqrt{\rho^2+\lambda^{-2}}}\right)^{n} =  \frac{1}{n-2} \lambda^{(n-2)}\xi^n\Pi^{n-1}_0(i\nu) .
\end{split}
\end{equation}
For the sum of all such diagrams $W'(i\nu) = \sum_{n=2}^\infty W_n(i\nu)$, we thus obtain 
\begin{equation} \label{RPA W}
\begin{split}
    W'(i\nu) &= W_2(i\nu) + \xi^2\Pi_0(i\nu)\sum_{n=3}^\infty\frac{1}{n-2}\big[\lambda^{-1}/\xi\Pi_0(i\nu) \big]^{-(n-2)}\\
    &= \xi^2\Pi_0(i\nu)\left[-\gamma/2-\log(\lambda^{-1} -\xi\Pi_0(i\nu)) + \mathcal{O}(\lambda^{-2})\right].
\end{split}
\end{equation}
Since $\Pi_0 <0$, we have a logarithmic singularity on the negative real axis at  $\xi_s = -1/(\lambda|\Pi_0|)$. The bare series diverges at the physical coupling $\xi = 1$ when the natural screening length $\propto \Pi_0^{-1}$ is comparable to the artificial screening length $\lambda$. However a resummation of the series for the $\log$ function beyond the convergence radius $|\xi_s|$ is straightforward and provides access to the Coulomb regime in which $\Pi^{-1} \ll \lambda$, with a residual linear dependence on $\lambda^{-1} \ll 1$ which can be extrapolated to $\lambda^{-1} =0$.

\begin{figure}[h!]
\centering
\begin{minipage}{0.48\linewidth}
    \centering
    \includegraphics[width=\linewidth]{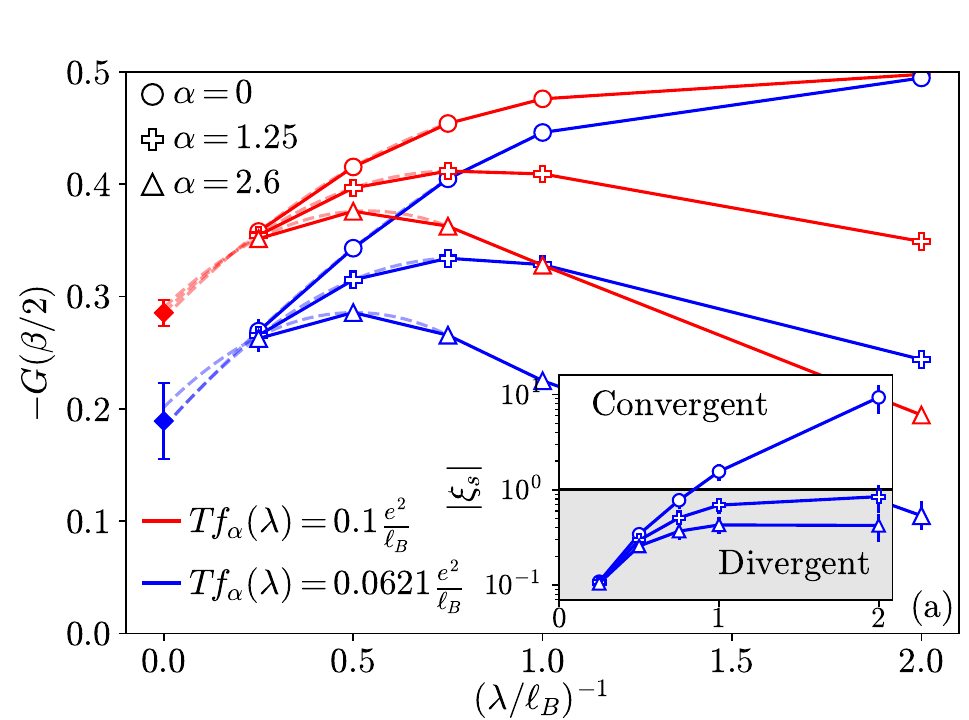}
\end{minipage}
\hfill
\begin{minipage}{0.48\linewidth}
    \centering
    \includegraphics[width=\linewidth]{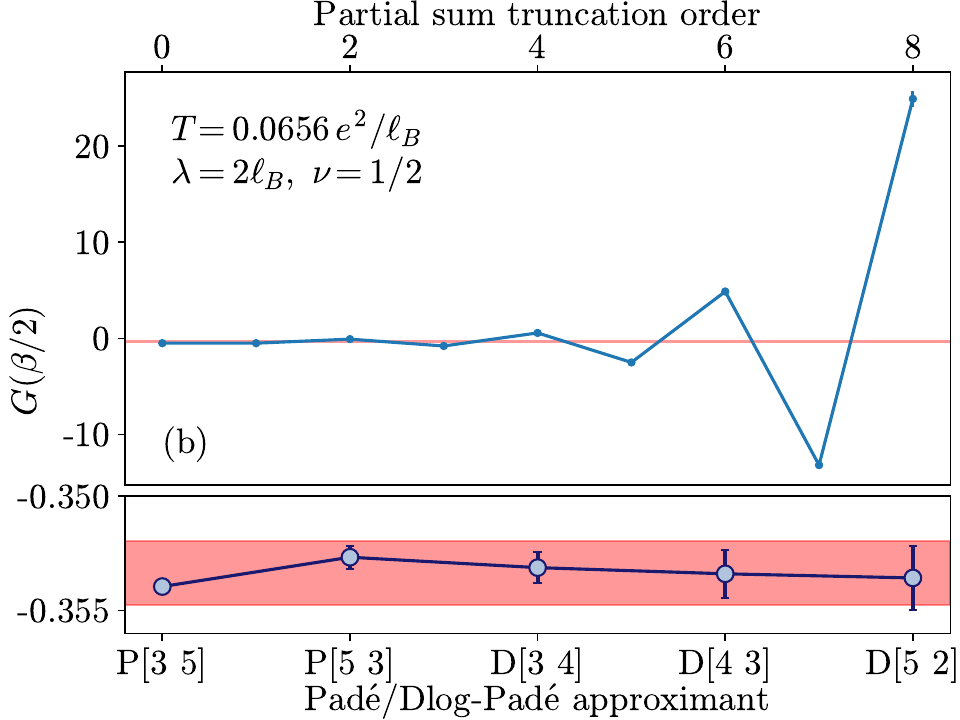}
\end{minipage}
\caption{(a) Example of $\lambda$-dependence of the Green's function $G(\tau=\beta/2)$ and it's extrapolation to the Coulomb limit $\lambda^{-1} \to 0$. Data for $\nu=1/2$ and two temperatures---$T=0.1e^2/l_B$ and $T=0.0621e^2/l_B$---approached in the Coulomb limit, $\lambda \to \infty$, are shown. The emergence of the linear regime---as predicted by Eq.\eqref{RPA W}---is seen at small $\lambda^{-1}$ where the series is strongly divergent. The convergence radius (equal to the absolute value of the singularity $\xi_s$ nearest the origin) for each $\lambda$ is shown in the inset for the lower temperature curves $Tf_\alpha(\lambda)=0.0621e^2/\ell_B$. The series at the physical coupling $\xi=1$ diverges in the shaded region. (b) Resummation of the Green's function $G(\tau=\beta/2)$ at temperature $T=0.0656 e^2/\ell_B$ and $\lambda = 2\ell_B$ and at half-filling $\nu=1/2$. Upper panel: The divergent partial sums of the series; the horizontal line is the result of the resummation. Lower panel: the corresponding results of the Pad\'e (P) and Dlog-Pad\'e (D) extrapolations for different choices of the orders $[p,q]$. The spread in their values, indicated by the shaded horizontal band, gives our conservative estimate of the systematic error of the extrapolation.}
\label{extrapolations}
\end{figure}

The $\lambda$-dependence of our numerically-exact data for the spectral function proxy $-\beta G(\tau=\beta/2)$ is shown at half filling in Fig.\ref{extrapolations}(a). We find that at small $\lambda$, where the series is strongly divergent, the reconstructed value exhibits an asymptotically linear behaviour (consistent with Eq.\eqref{RPA W}), which is straightforwardly extrapolated to the Coulomb limit $\lambda\to \infty$. The robustness of the extrapolation can be further verified by performing extrapolations along different paths in the $(\lambda , T)$ parameter space, with all paths constructed to meet at the Coulomb point $(\lambda^{-1} = 0, T)$ for a given temperature $T$. For instance, the most natural choice is to extrapolate with respect to $\lambda$ at fixed temperature, but one could also consider a path with a $\lambda$-dependent rescaling of the temperature $(\lambda, f_\alpha(\lambda)\times T)$ where $f_\alpha(\lambda)$ is a family of rescaling functions parametrised by $\alpha$ with $f_\alpha(\lambda^{-1} =0) = 1$. Motivated by the fact that the Coulomb energy scale is asymptotic to $\lambda^{-3}$ for $\lambda \to 0$ while it is finite for $\lambda \to \infty$, we make the choice $f_\alpha(\lambda) = 1+\alpha \lambda^{-3}$. The results of the extrapolation along such $\alpha$-dependent paths, shown in Fig. \ref{extrapolations}(a), are independent of $\alpha$ and agree within the errorbars, showing the robustness of the extrapolation.
The curves for the temperature dependence of $\tilde\rho=-\beta G(\tau=\beta/2)$ shown in Fig. \ref{fig:beta_over_2}(c) are obtained in this way, and the final error is obtained as a combined estimate from a few reasonable choices of $\alpha$ (the same as those in Fig. \ref{extrapolations}(a)).

While the example Eqs.~\eqref{W_n}-\eqref{RPA W} focussed on the 2D electron gas in a strong magnetic field, a similar analysis holds for the 3D electron gas in zero external field. In this case, the same calculation yields a square-root singularity located at $\xi_s = -1/(\lambda^2|\Pi_0|)$. Therefore analytic continuation beyond the singularity at finite $\lambda$ provides a general method for accessing Coulomb physics by a Feynman expansion in the bare potential, combined with an extrapolation to $\lambda =\infty$.

\section{Resummation of the series} \label{Pade appendix}

Within the diagrammatic Monte Carlo approach, an arbitrary physical observable $A(\xi)$ is computed as a function of the coupling strength $\xi$ by construction of its Taylor series $A_N(\xi) = \sum_{n=0}^N a_n \xi^n$ up to a maximal diagram order $N$. Using only the coefficients $a_n$, which are calculated numerically exactly up to order $N$, the task then is to evaluate $A(\xi)$ by an extrapolation of the series to infinite diagram order.

In the strongly correlated regimes considered in the main text, the  diagrammatic series for the Green's function is either slowly convergent or divergent, depending on the temperature and the screening range $\lambda$. Therefore, a resummation procedure is required to perform the extrapolation to infinite diagram order and thus extract the value of the function $A(\xi)$ behind its series $\sum_{n} a_n\xi^n$. For this we rely on the established techniques of Pad\'e and Dlog-Pad\'e resummation \cite{Baker1961}. Such methods have been tested and used to obtain controlled results in systematic expansions of diverse types since the 1960s studies of the Ising model~\cite{Baker1961}, and they continue to be widely applied, including in high-temperature expansions~\cite{Domb1974}, numerical linked-cluster expansions~\cite{Rigol2006}, as well as diagrammatic Monte Carlo methods~\cite{Simkovic2019, Simkovic2020, Schafer2021}. They involve the construction of approximants that are designed to capture different possible analytic structures of $A(\xi)$: Pad\'e approximants represent the function as a sum of simple poles, while Dlog-Pad\'e approximants describe it as a product of algebraic singularities.

Specifically, Pad\'e resummation allows one to assign a value to $A(\xi)$ by means of the construction of a rational function $A^{\mathrm{Pad\acute{e}}}_{[p,q]}(\xi)$ that has precisely the same Taylor series coefficients as $A$ up to order $N$, where $p,q$ are the orders of the polynomials in the numerator and denominator respectively, and which satisfy $p+q = N$. Similarly, the Dlog-Pad\'e method constructs the approximating function $A^{\mathrm{Dlog}}_{[p,q]}(\xi) = A(0)\exp\left(\int_0^\xi d\xi'~\tilde{A}_{[p,q]}(\xi')\right)$, where $\tilde{A}_{[p,q]}$ is the $[p,q]$ Pad\'e approximant for the derivative of the logarithm of $A(\xi)$, while $p+q = N-1$. Both types of approximating function allow the evaluation of the function $A(\xi)$ at arbitrary values of $\xi$, including beyond the convergence radius where the Taylor series for $A$ is divergent, by taking the limit $\lim_{N\to\infty} A_{[p,q]}(\xi) = A(\xi)$. 

Although, in general, an individual approximant $A_{[p,q]}(\xi)$ (either Pad\'e or Dlog-Pad\'e) is not expected to capture the unknown analytic structure of $A(\xi)$ exactly, at sufficiently large truncation order $N$ this discrepancy may not be resolvable beyond the statistical error in $A_{[p,q]}(\xi)$ propagated from the coefficients $a_n$.
Crucially, if, for a given $\xi$, a mismatch between different approximants can be resolved, it provides an internal estimate of the systematic error of the extrapolation: since the physical result $A(\xi)$ does not depend on the free parameters $[p,q]$ of its approximating function, the variation of the reconstructed value under changes of these parameters offers a reliable measure of the systematic uncertainty. 

For computing the filling fraction, we employ an additional modification of the resummation procedure so that the particle-hole symmetry is explicitly enforced. Specifically, due to the projection onto the LLL, the filling fraction must satisfy $\nu(\tilde{\mu}) = 1-\nu(-\tilde{\mu})$ for arbitrary $\xi$, where $\tilde{\mu}$ is the Hartree-Fock shifted chemical potential defined in Eq.~\eqref{HF GF}. The coefficients of the expansion at order $n$ thus satisfy $\nu_n(\tilde{\mu}) = -\nu_n(-\tilde{\mu})$ for $n\geq1$, while for the zeroth order $\nu_0(\tilde{\mu}) = 1-\nu_0(-\tilde{\mu})$.
We see from the above that any spurious particle-hole symmetry breaking in the Pad\'e and Dlog-Pad\'e approximants will arise purely from the zeroth order coefficient, since this is the only part of the series that does not simply pick up a minus sign under the particle-hole transformation. Therefore, to restore particle-hole symmetry, we extrapolate the series with an arbitrary zeroth-order coefficient, whose value we choose by demanding that the extrapolated value depends least on the choice of approximating function (i.e. the values of $[p,q]$). Not only does this enforce the correct symmetry, but by construction it also reduces the discrepancy between the different approximating functions, implying that the true analytic structure of the function is more accurately captured.

As a concrete example, in Fig.\ref{extrapolations}(b) we show an extrapolation of the series for $G(\tau=\beta/2)$ in the longer-range case $\lambda = 2\ell_B$. The partial sum (top panel) shows a strong divergence, owing to the Coulomb singularity on the negative real axis, while the $N=8$ Pad\'e and Dlog-Pad\'e approximants yield consistent estimates. The spread in their values, including their statistical errors---shown by the shaded horizontal band---provides our conservative estimate of the resummation error bars used throughout this work.

\end{document}